\documentclass[twoside,fleqn]{ActaStyle}
\usepackage{times,cite}

\usepackage{lscape}             
\usepackage{graphicx,epsfig}    
\usepackage[tbtags]{amsmath}    

\def\authorheading{J.~Stre\v{c}ka and M.~Ja\v{s}\v{c}ur}

\def\shorttitle{Disordered and ordered states...}

\begin{document}

\pagerange{1}{6}

\title{DISORDERED AND ORDERED STATES OF EXACTLY SOLVABLE 
ISING-HEISENBERG PLANAR MODELS WITH A SPATIAL ANISOTROPY\footnote{Presented 
at 15-th Conference of Czech and Slovak Physicist, Ko\v{s}ice, Slovakia, September 5-8, 2005}}

\author{J.~Stre\v{c}ka\email{jozkos@pobox.sk} and M.~Ja\v{s}\v{c}ur}
{Department of Theoretical Physics and Astrophysics, Faculty of Science, \\ 
P. J. \v{S}af\'{a}rik University, Park Angelinum 9, 040 01 Ko\v{s}ice, Slovakia}

\day{August 15, 2005}

\abstract{Ground-state and finite-temperature properties of a special class of exactly solvable
Ising-Heisenberg planar models are examined using the generalized decoration-iteration 
and star-triangle mapping transformations. The investigated spin systems exhibit an interesting 
quantum behaviour manifested in a remarkable geometric spin frustration, which appears notwithstanding 
the purely ferromagnetic interactions of the considered model systems. This kind of spin frustration
originates from an easy-plane anisotropy in the XXZ Heisenberg interaction between nearest-neighbouring 
spins that favours ferromagnetic ordering of their transverse components, whereas their 
longitudinal components are aligned antiferomagnetically.}


\section{Introduction}
\label{sec:intr} \setcounter{section}{1}\setcounter{equation}{0}

Quantum behaviour of low-dimensional solids belongs to the most fascinating topics 
to be ever discovered in the field of condensed matter physics. Generally, it is  
expected that the clearest manifestations of quantum phenomena should emerge in the 
low-dimensional antiferromagnets which consist of spin carriers having low quantum spin 
number. With regard to this, much effort has recently been devoted to the two-dimensional 
antiferromagnetic spin systems that have an obvious relevance in elucidating 
the high-$T_c$ superconductivity of cuprates \cite{Man91}, quantum phase transitions 
\cite{Sac99}, geometric spin frustration \cite{Ram01}, etc. On the other hand, the 
two-dimensional quantum ferromagnets remain rather unexplored from this point of view 
as their macroscopic quantum manifestations would be, on the contrary, rather surprising. 

Nevertheless, the two-dimensional ferromagnetic spin systems with a spatial anisotropy represent
very interesting systems as they may exhibit spontaneous long-range order possibly affected
by quantum effects. A strong evidence supporting this statement has been provided by a recent
exploration of outstanding quantum antiferromagnetic phase, which was discovered in the 
mixed-bond Ising-Heisenberg models (to be further abbreviated as IHMs) despite the pure 
ferromagnetic interactions of the considered model systems \cite{Str02, Str03, Str04}. This 
remarkable finding suggests that an interesting quantum behaviour can also be found in the anisotropic ferromagnetic systems owing to a competition between the easy-plane XXZ Heisenberg interactions 
and the familiar Ising interactions of the easy-axis type. In the present article we shall 
show that this unusual behaviour represents a generic feature of wide class of the mixed-bond 
IHMs that arises independently of the lattice topology and hence, on close-packed lattices it 
leads to disordered ground state due to a presence of peculiar geometric spin frustration. 

This paper is organized as follows. In the next section we shall briefly describe model 
systems and the method that enables to obtain their exact solution. The most interesting results 
are presented and discussed in Section \ref{sec:result}. Finally, some concluding remarks are 
given in Section \ref{sec:conc}.
        
\section{Model systems and their exact solution}
\label{sec:model}

Let us begin by considering periodic planar lattices composed of two kinds of spin sites; so-called 
Ising spins $\mu$ which interact with other spins through extremely anisotropic coupling $J_{\rm I}$ containing their one spatial component only and the Heisenberg spins $S$ which interact among themselves 
via spatially anisotropic coupling $J(\Delta)$ containing all three spatial components. The parameter 
$\Delta$ allows to control a spatial anisotropy in the XXZ Heisenberg interaction, which can be either 
of easy-axis ($\Delta < 1$) or easy-plane ($\Delta > 1$) type. The total Hamiltonian of mixed-bond 
IHMs then includes both Ising as well as Heisenberg bonds and thus, it acquires this simple form
\begin{equation}
\hat {\cal H} = - J \sum_{(i,j)} [ \Delta (\hat S_i^x \hat S_j^x + \hat S_i^y \hat S_j^y) + 
               \hat S_i^z \hat S_j^z ] - J_{\rm I} \sum_{(k, l)} \hat S_k^z \hat \mu_l^z,
\label{eq:1}
\end{equation}
where $\hat S_i^{\alpha}$ ($\alpha = x,y,z$) and $\hat \mu_j^z$ are spatial components of spin-1/2 operators, the first summation is carried out over the nearest-neighbouring Heisenberg spin pairs and the second summation runs over the nearest-neighbouring pairs of the Ising and Heisenberg spins, respectively.

\begin{figure}[t]
\begin{center}
\includegraphics[width=3.5in]{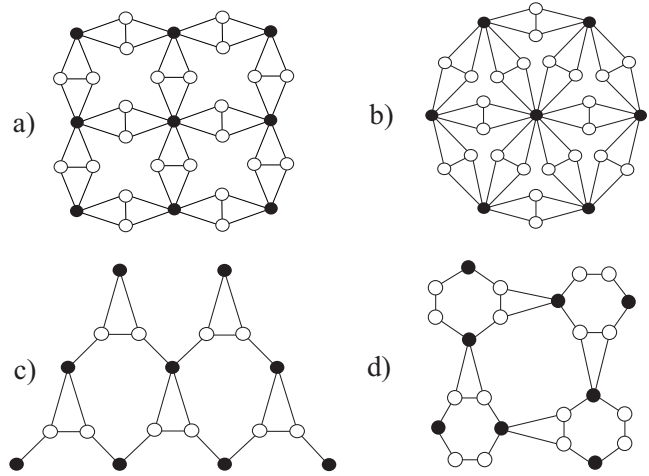} \\
\end{center}
\caption{Portions of several Ising-Heisenberg lattices. Full (empty) circles denote lattice 
positions of the Ising (Heisenberg) spins. Lattices a) and b) are further treated as a model A,
lattices c) and d) as a model B.}
\label{fig:1}
\end{figure}

\begin{figure}[t]
\begin{center}
\includegraphics[width=4.5in]{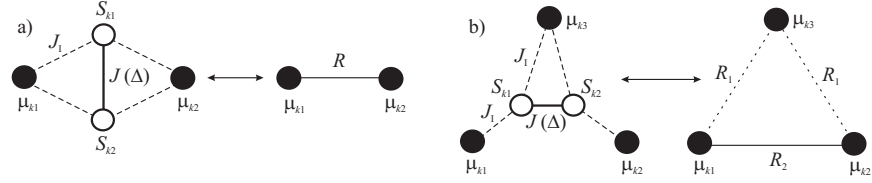} \\
\end{center}
\caption{Schematic representation of an extended: a) decoration-iteration and 
b) star-triangle transformation.}
\label{fig:2}
\end{figure}

In what follows we shall restrict ourselves to a special sub-class of the model Hamiltonian (\ref{eq:1}), namely, we will further refer just to spin systems that consists of Heisenberg spin pairs each
interacting with either two or three nearest-neighbouring Ising spins. Several examples of such lattices are sketched in Fig.~\ref{fig:1}, where full (empty) circles denote lattice positions of the Ising (Heisenberg) spins. A common feature of the lattices a) and b) is that the Heisenberg spin pairs are placed in between two outer Ising spins (model A), while on lattices c) and d) they are surrounded by three outer Ising spins 
(model B). In view of further manipulations, it is very advisable to rewrite the total Hamiltonian as a 
sum of bond Hamiltonians, i.e. $\hat {\cal H} = \sum_{k} \hat {\cal H}_k$, each containing all the interaction 
terms of one Heisenberg pair (spin clusters on lhs of Fig.~\ref{fig:2})
\begin{equation}
\hat {\cal H}_{k} = - J [\Delta (\hat S_{k1}^x \hat S_{k2}^x + \hat S_{k1}^y \hat S_{k2}^y) + 
               \hat S_{k1}^z \hat S_{k2}^z \} - \hat S_k^z h_1 - \hat S_k^z h_2.
\label{eq:2}
\end{equation}
Above, $h_1 = h_2 = J_{\rm I} (\hat \mu_{k1}^z + \hat \mu_{k2}^z)$ for model A, while
$h_1 = J_{\rm I} (\hat \mu_{k1}^z + \hat \mu_{k3}^z), h_2 = J_{\rm I} (\hat \mu_{k2}^z + \hat \mu_{k3}^z)$ 
for model B. On account of the commutability between different bond Hamiltonians, the partition function 
of IHMs can be partially factorized into a product of bond partition functions 
\begin{equation}
{\cal Z}_{IHM} = \sum_{\{ \mu \}} \prod_{k = 1}^{N} \mbox{Tr}_k \exp(- \beta \hat {\cal H}_k)
               = \sum_{\{ \mu \}} \prod_{k = 1}^{N} {\cal Z}_{k},
\label{eq:3}
\end{equation}
where $\beta = 1/(k_{\rm B} T)$, $k_{\rm B}$ is Boltzmann's constant, $T$ denotes absolute temperature,
the symbol $\mbox{Tr}_k$ stands for a trace over spin degrees of freedom of $k$th Heisenberg 
spin pair and the summation $\sum_{\{ \mu \}}$ runs over all available configurations of the Ising spins.

A crucial step of our approach represents calculation of the bond partition function ${\cal Z}_{k}$ 
accomplished by adopting the same steps as used in Ref. \cite{Str02} and its subsequent replacement 
by suitable mapping transformation. For this purpose, we shall utilize generalized 
versions of the decoration-iteration \cite{Syo51} and star-triangle \cite{Ons44} transformations
originally introduced by solving planar Ising models. It is noteworthy, however, that 
the mapping transformations have been later on remarkably generalized by Fisher \cite{Fis59, Syo72} 
who firstly pointed out that {\it decorating} Ising spins may be, in principle, substituted by arbitrary 
statistical systems. In our case, both the transformations can be schematically illustrated as
in Fig.~\ref{fig:2} and mathematically they can be expressed as follows
\begin{eqnarray}
{\cal Z}_k \! \! \! \! \! &=& \! \! \! \! \! 2 \exp(\beta J/4) \cosh(\beta h_1) + 
2 \exp(- \beta J/4) \cosh(\beta J \Delta/2) = A \exp(\beta R \mu_{k1}^z \mu_{k2}^z), 
\label{eq:4} \\
{\cal Z}_k \! \! \! \! \! &=& \! \! \! \! \! 2 \exp(\beta J/4) \cosh[\beta (h_1 + h_2)/2] +
2 \exp(- \beta J/4) \cosh[\beta \sqrt{(h_1 - h_2)^2 + (J \Delta)^2}/2] \nonumber \\
\! \! \! \! \! &=& \! \! \! \! \! 
B \exp[\beta R_1 (\mu_{k1}^z \mu_{k3}^z + \mu_{k2}^z \mu_{k3}^z) + \beta R_2 \mu_{k1}^z \mu_{k2}^z],
\label{eq:5} 
\end{eqnarray}
where $A$ and $R$ ($B$, $R_1$, and $R_2$) are unknown mapping parameters emerging in the 
decoration-iteration (star-triangle) transformation that are unambiguously given by a self-consistency 
condition of the mapping relations. Unknown mapping parameters can be calculated following the standard procedure \cite{Fis59, Syo72}, for instance, the mapping (\ref{eq:4}) holds for model A if and only if
\begin{equation}
A = 2 \exp(\beta J/4)(V_1 V_2)^{1/2}, \quad \beta R = 2 \ln(V_1/V_2),  
\label{eq:6}
\end{equation}
and the functions $V_1$ and $V_2$ are defined as follows
\begin{eqnarray}
V_1 \! \! \! \! \! &=& \! \! \! \! \! \cosh(\beta J_{\rm I}) + \exp(- \beta J/2) \cosh(\beta J \Delta/2),
\label{eq:7} \\
V_2 \! \! \! \! \! &=& \! \! \! \! \! 1 + \exp(- \beta J/2) \cosh(\beta J \Delta/2).  
\label{eq:8}
\end{eqnarray}
One the other hand, the model B can be treated within the mapping transformation (\ref{eq:5}) when
\begin{equation}
B = 2 \exp(\beta J/4)(V_1 V_2 V_3^2)^{1/4}, 
\quad \beta R_1 = \ln(V_1/V_2), 
\quad \beta R_2 = \ln(V_1 V_2/V_3^2). 
\label{eq:9}
\end{equation}
In addition to the functions $V_1$ and $V_2$ defined above we have introduced here another function
\begin{eqnarray}
V_3 = \cosh(\beta J_{\rm I}/2) + \exp(- \beta J/2) 
      \cosh \Bigl(\beta \sqrt{J_{\rm I}^2 + (J \Delta)^2}/2 \Bigr).  
\label{eq:10}
\end{eqnarray} 

At this stage, a direct substitution of the transformations (\ref{eq:4}) and (\ref{eq:5}) into the formula (\ref{eq:3}) gained for the partition function of IHMs leads to a mapping relationship with partition functions of the spin-1/2 Ising models on undecorated lattices (scheme from Fig.~\ref{fig:2}), in fact, 
\begin{eqnarray}
\mbox{Model A:} && \quad {\cal Z}_{IHM} = A^N {\cal Z}_{IM} (\beta, R),
\label{eq:11} \\
\mbox{Model B:} &&  \quad {\cal Z}_{IHM} = B^N {\cal Z}_{IM} (\beta, R_1, R_2).
\label{eq:12}
\end{eqnarray}
Notice that the mapping relations (\ref{eq:11}) and (\ref{eq:12}) allow simple calculation of other quantities as well. For easy reference, we list the most important ones: sub-lattice magnetization of the Ising spins $m_{i}^{z} \equiv \langle \hat \mu_{k1}^z \rangle$, sub-lattice magnetization of the Heisenberg spins $m_{h}^{z} \equiv \langle \hat S_{k1}^z \rangle$, pair correlation functions between the nearest-neighbouring Heisenberg spins $C_{hh}^{xx} \equiv \langle \hat S_{k1}^x \hat S_{k2}^x \rangle$, 
$C_{hh}^{zz} \equiv \langle \hat S_{k1}^z \hat S_{k2}^z \rangle$, and between the nearest-neighbouring 
Ising and Heisenberg spins $C_{ih}^{zz} \equiv \langle \hat \mu_{k1}^z \hat S_{k1}^z \rangle$.

\section{Results and Discussion}
\label{sec:result}

In this section, the most interesting results for both investigated model systems will be presented and discussed in detail. Although the results obtained in the preceding section hold irrespectively of 
a character of the exchange parameters $J$ and $J_{\rm I}$, our further analysis will be for simplicity restricted to the ferromagnetic systems ($J > 0$, $J_{\rm I} > 0$) only. In order to avoid confusion, 
the results obtained for both investigated model systems will be discussed separately.

\subsection{Model A}
\label{subsec:A}

Initially, let us look more closely on the ground-state behaviour. The first-order transition line 
$\Delta_c = 1 + 2 J_1/J$ separates regions of stability of two different ground-state phases for both 
IHMs with a symmetric decoration shown in Fig.~\ref{fig:1}a)-b). As could be expected, there 
occurs a simple {\em ferromagnetic phase} (FP) characterized by saturated sub-lattice magnetization 
$m_i = 1/2$, $m_h = 1/2$, as well as pair correlation functions $C_{hh}^{zz} = 1/4$, $C_{ih}^{zz} = 1/4$ 
when the exchange anisotropy is below its critical value $\Delta_c$. Within FP one also finds 
$C_{hh}^{xx} = 0$ clearly indicating classical character of this long-range ordered phase. 
However, there also appears a remarkable {\em disordered phase} (DP) without spontaneous ordering ($m_i = 0$, 
$m_h = 0$) when the exchange anisotropy exceeds its critical value $\Delta_c$. It should be emphasized, nevertheless, that the Heisenberg spin pairs remain strongly correlated even within DP as suggested by $C_{hh}^{xx} = 1/4$ and $C_{hh}^{zz} = -1/4$. By contrast, the nearest-neighbouring Ising and Heisenberg spins are completely uncorrelated, indeed, one finds $C_{ih}^{zz} = 0$ within DP. Thus, it can be readily understood that DP emerges due to a peculiar geometric spin frustration originating from an antiferromagnetic alignment 
of the longitudinal components of the Heisenberg spins that accompanies preferred ferromagnetic 
ordering of their transverse components. In agreement with this statement, it can be easily checked 
that the function $|\Psi_{{\rm DP}} \rangle = \prod_{i} | \! \! \updownarrow \rangle_i 
\prod_{k} \frac{1}{\sqrt{2}} (|\! \! \uparrow \downarrow \rangle + | \! \! \downarrow \uparrow \rangle)_k$
describes the ground state of DP (the first product is carried out over all Ising spins and the second one runs over all Heisenberg spin pairs). According to this, one may conclude that all Ising spins are free to flip in DP on account of the spin frustration caused by Heisenberg spin pairs, which exhibit an interesting coexistence of ferromagnetic (transverse) and antiferromagnetic (longitudinal) correlations.  

\begin{figure}[t]
\begin{minipage}[t]{0.48\linewidth}
\begin{center}
\includegraphics[width=2.5in]{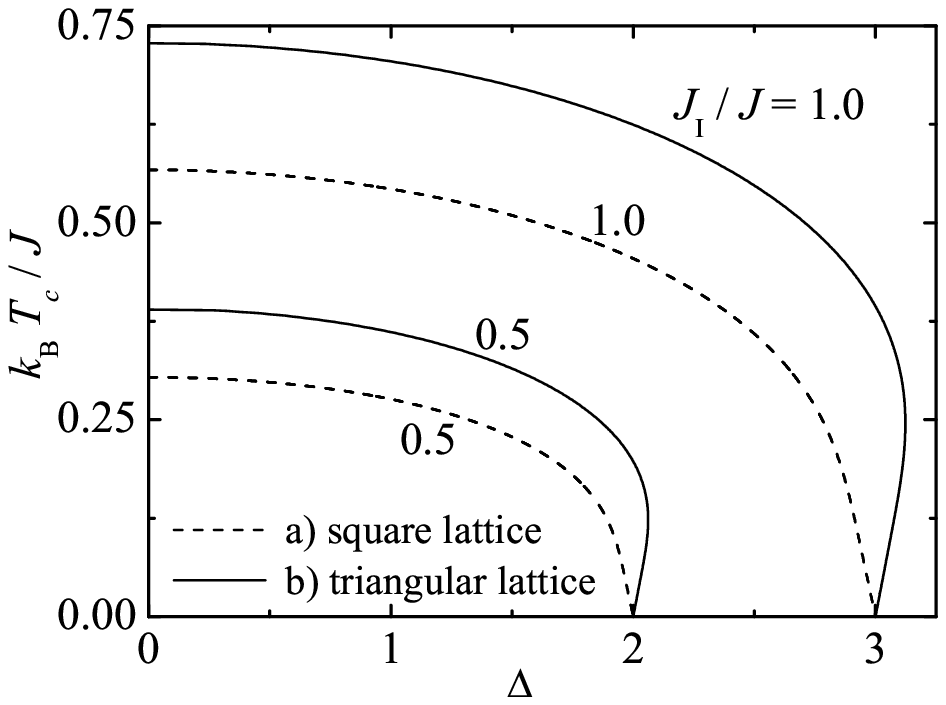} \\
\end{center}
\caption{Critical temperatures of IHMs with a symmetric decoration 
of the square (broken lines) and triangular (solid lines) lattices 
shown in Fig.~\ref{fig:1}a)-b).}
\label{fig:3}
\end{minipage}%
\hspace{0.04\textwidth}%
\begin{minipage}[t]{0.48\linewidth}
\begin{center}
\includegraphics[width=2.5in]{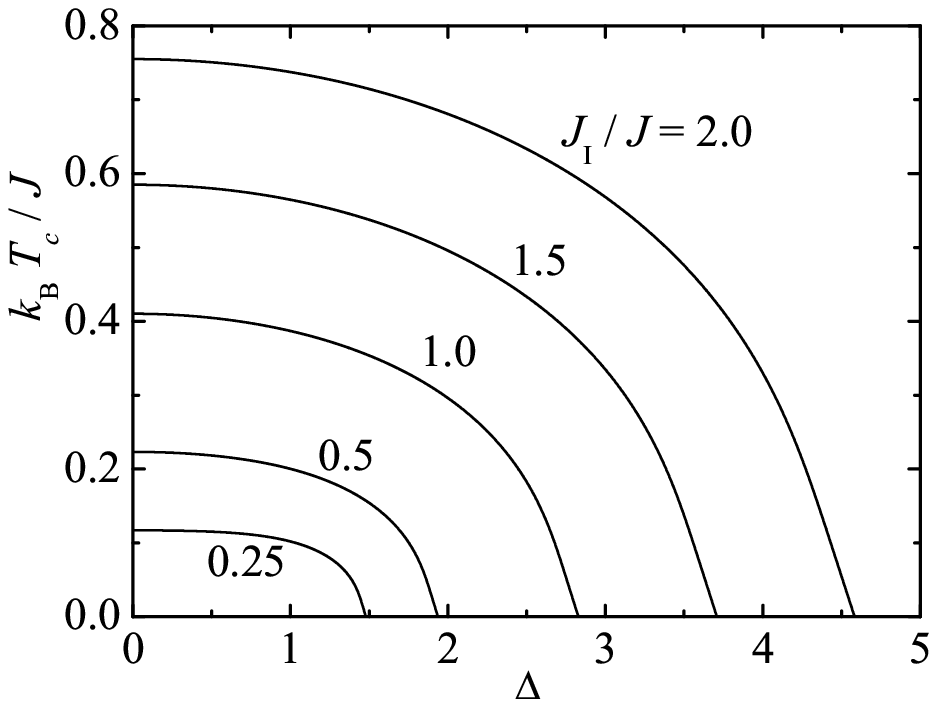} \\
\end{center}
\caption{Critical temperatures of IHMs with an asymmetric decoration of the 
triangular (Fig.~\ref{fig:1}c) and orthogonal-dimer (Fig.~\ref{fig:1}d) lattices.}
\label{fig:4}
\end{minipage}
\end{figure}

Now, let us step forward to the finite-temperature behaviour. The variation of critical 
temperature of IHMs with a symmetric decoration of the square (Fig.~\ref{fig:1}a) and triangular 
(Fig.~\ref{fig:1}b) lattices is displayed in Fig.~\ref{fig:3} for two different values of $J_{\rm I}/J = 0.5$ 
and $1.0$. As one can see, the critical lines of both the lattices terminate at common disorder point 
as predicted by the ground-state analysis. The most obvious difference in the criticality 
of both IHMs thus rests in a reentrant transition of the triangular lattice (Fig.~\ref{fig:1}b) 
observable in the vicinity of disorder points. 
  
\subsection{Model B}
\label{subsec:B}

Ground-state behaviour of IHMs with an asymmetric decoration of the triangular (Fig.~\ref{fig:1}c) and orthogonal-dimer (Fig.~\ref{fig:1}d) lattices is strongly reminiscent to the one discussed earlier. 
As a matter of fact, the ground state constitute the aforementioned FP and DP that are bounded by the 
first-order transition line $\Delta_c = \sqrt{(1 + 3 J_{\rm I}/J)(1 + J_{\rm I}/J)}$. 
The occurrence of DP can be once again related to the spin frustration caused by the Heisenberg 
spins, which are antiferromagnetically aligned in the longitudinal direction and ferromagnetically
coupled in the transverse direction. In the consequence of that, all Ising spins must be strongly 
frustrated because they interact with the longitudinal components of the Heisenberg spins only. 
Hence, one may conclude that an appearance of the disordered state
$|\Psi_{{\rm DP}} \rangle = \prod_{i} | \! \! \updownarrow \rangle_i 
\prod_{k} \frac{1}{\sqrt{2}} (|\! \! \uparrow \downarrow \rangle + | \! \! \downarrow \uparrow \rangle)_k$ above certain $\Delta_c$ (depending on the lattice topology) represents a generic feature of all IHMs having 
the magnetic structure of some close-packed lattice.

Finally, we shall close our discussion with the finite-temperature phase diagram of IHMs with the asymmetric decoration of the triangular and orthogonal-dimer lattices. It is quite interesting to ascertain from 
Fig.~\ref{fig:4} that the critical temperatures of both these lattices overlap as they exhibit 
almost the same decline (not discernible within the displayed scale) upon strengthening of $\Delta$.

\section{Conclusion}
\label{sec:conc}
In the present article we have provided an exact solution to a special sub-class of IHMs 
which consist of the Heisenberg spin pairs coupled either to two or three outer Ising spins. 
The investigated spin systems exhibit an interesting quantum behaviour manifested in a remarkable
kind of geometric spin frustration, which appears notwithstanding the purely ferromagnetic 
interactions of the considered model systems. We found a convincing evidence that this behaviour 
emerges due to a competition between the familiar Ising interactions and the easy-plane XXZ Heisenberg interactions, which favour ferromagnetic (antiferromagnetic) ordering of the transverse (longitudinal) components of the Heisenberg spin pairs. 
The most important question that arises from our study is whether this behaviour represents a general feature 
of all quantum models with mixed easy-axis and easy-plane bonds and whether it would persist also in 
the pure Heisenberg spin systems. This question represent a great challenge for our next investigations.

\begin{ack}
The authors acknowledge financial support provided by Ministry of Education 
of Slovak Republic given under the grants VEGA 1/2009/05 and APVT 20-005204.
\end{ack}

\end{document}